# DSL-based Design Space Exploration for Temporal and Spatial Parallelism of Custom Stream Computing

Kentaro Sano
Graduate School of Information Sciences,Tohoku University
6-6-01 Aramaki Aza Aoba, Sendai 980-8579, JAPAN
Email: kentah@caero.mech.tohoku.ac.jp

*Abstract*—Stream computation is one of the approaches suitable for FPGA-based custom computing due to its high throughput capability brought by pipelining with regular memory access. To increase performance of iterative stream computation, we can exploit both temporal and spatial parallelism by deepening and duplicating pipelines, respectively. However, the performance is constrained by several factors including available hardware resources on FPGA, an external memory bandwidth, and utilization of pipeline stages, and therefore we need to find the best mix of the different parallelism to achieve the highest performance per power. In this paper, we present a domain-specific language (DSL) based design space exploration for temporally and/or spatially parallel stream computation with FPGA. We define a DSL where we can easily design a hierarchical structure of parallel stream computation with abstract description of computation. For iterative stream computation of fluid dynamics simulation, we design hardware structures with a different mix of the temporal and spatial parallelism. By measuring the performance and the power consumption, we find the best among them.

## I. INTRODUCTION

Recently, FPGA-based custom computing has been attracting a lot of application developers especially in the big-data and supercomputing fields, where not only performance but also power consumption is a very important. Since it is difficult to further increase a clock frequency of a general-purpose microprocessor, so far many-core accelerators such as GPUs have been considered as a promising solution to obtain higher computing performance. However, achievable performance is limited by the overall power budget of an entire system, and therefore power efficiency is considered as a key to large-scale computation.

On the other hand, custom computing with FPGAs is expected to provide comparable performance at much lower power consumption. Custom circuits are able to effectively achieve high performance by exploiting spatial and temporal parallelism of computing problems at a low clock frequency. Moreover, recent advancement of FPGAs fabricated by cutting-edge semiconductor technologies is bringing high potential for efficient and high performance computation due to on-chip integration of many hard macros such as block RAMs, high-speed I/O blocks, and DSP blocks. Especially emerging state-of-the-art FPGA devices are capable of very high performance numerical computation at a low power with their hard floating-point DSP blocks [7].

Stream computing is one of the promising approaches for efficient computation with custom hardware. This is because 1) deep pipelines can increase the number of operations performed per memory access, and 2) regular accesses for streaming data fully utilize a precious bandwidth of external memories. In addition, dedicated hardware designs bring efficient utilization of resources on FPGAs by adaptively giving a various mix of different operators and functions, including an adder, a multiplier, a divider, and a square root function. So far, researches have been reported on their successful high-performance stream computing with FPGAs [6], [9].

However, productivity still remains as a big issue not only in designing custom hardware, but also in exploring design space to obtain the best performance per power. In the case of stream computation, we can exploit the two types of parallelism: spatial, and temporal. By duplicating a pipeline to exploit the spatial parallelism, we can increase operations performed at every cycle, resulting in higher performance until all the available hardware resources are consumed. However, this also increases bandwidth requirements to an external memory, and the scalability is limited by the available bandwidth. On the other hand, by deepening a pipeline to exploit the temporal parallelism, we can increase operations per memory access, resulting in higher performance with the same memory bandwidth. However, too long pipelines suffer from low utilization due to the prologue and epilogue effects in pipelining. Thus, the best mix of the two different parallelism depends on these constraints, and therefore we have to find the optimal one for individual application.

In this paper, we present a domain-specific language (DSL) based approach to easily explore a design space for spatially and/or temporally parallel stream computation with FPGA. Our own DSL, called a stream processing description (SPD), allows us to intuitively describe formulae and submodule calls for various computation and structures of custom hardware in a software-like abstraction level. In design exploration, we design various parallel-configurations in SPD to be compiled with our SPD compiler, and then find the best among them by evaluating performance and power consumption of their actually working implementations with FPGA.

So far, several languages are proposed for stream processing, including StreamIt [11], and its parallelism is studied [3]. There are also presented stream computing compilers targeting FPGAs, non-commercial ones [4], [5] and commercial ones [1], [8]. Our DSL is designed for compact and intuitive description of hierarchal and modular connection of hardware modules for explicit parallelism. In this study, we apply it to design space exploration for high-performance custom computing of the scientific numerical simulation. Contributions of this work are:

1) DSL, called SPD for custom stream computation,
2) Framework for DSL-based design space exploration,
3) Case study for FPGA-based fluid dynamics simulation.

The SPD compiler used in this work is an extended version of the previous one published in [10]. The extension was made

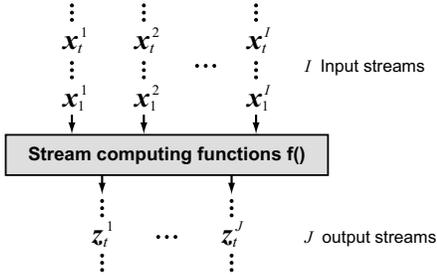

Fig. 1. Definition of stream computing.

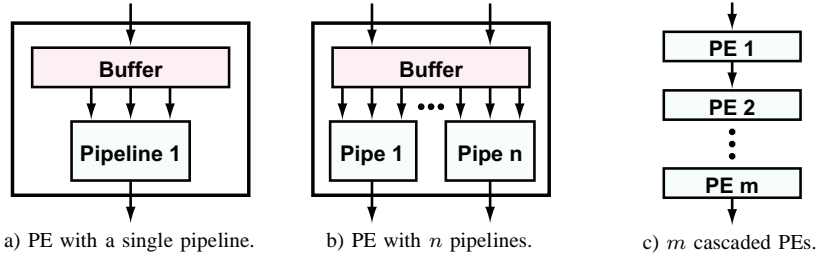

a) PE with a single pipeline.

b) PE with $n$ pipelines.

c) $m$ cascaded PEs.

Fig. 2. Processing element (PE) (a), spatial parallelism (b), and temporal parallelism (c).

essentially for hierarchical and modular description capability.

This paper is organized as follows. Section II describes parallelism of iterative stream computation, and SPD for design of stream-computing hardware. Section III gives a design of an application example and evaluation. Finally, Section IV gives conclusions and future work.

## II. DSL-BASED DESIGN SPACE EXPLORATION

### A. Stream computation

Here we define stream computation which is targeted by our domain-specific language (DSL). Stream computation has $I$ input data streams $\boldsymbol{x}^i$ ($1 \leq n \leq I$) and $J$ output data streams $\boldsymbol{z}^j$ ($1 \leq m \leq J$), each of which has elements as follows:

$$\boldsymbol{x}^i \equiv \{ \boldsymbol{x}^i_1, \boldsymbol{x}^i_2, ..., \boldsymbol{x}^i_t, ..., \boldsymbol{x}^i_T \}, \quad (1)$$
$$\boldsymbol{z}^j \equiv \{ \boldsymbol{z}^j_1, \boldsymbol{z}^j_2, ..., \boldsymbol{z}^j_t, ..., \boldsymbol{z}^j_T \}. \quad (2)$$

Each input or output stream has $T$ scalar elements, incoming or outcoming in order of a time $t = 1, 2, ..., T$. As shown in Fig.1, we model stream computation with a function $f^j()$ for the $j$-th output stream:

$$z^j_t = f^j \left( \{..., \boldsymbol{x}^1_t, ...\}, \{..., \boldsymbol{x}^2_t, ...\}, ..., \{..., \boldsymbol{x}^I_t, ...\} \right), \quad (3)$$

which means that the output of the $j$-th stream at a time $t$ is obtained by computing a given function with the input elements at $t$ and their offset ones if necessary. For example, stencil computation of a single variable with a $3 \times 3$ star stencil on a $256 \times 256$ grid can be streamed with

$$z_t = f \left( \boldsymbol{x}_{t-256}, \ \boldsymbol{x}_{t-1}, \ \boldsymbol{x}_t, \ \boldsymbol{x}_{t+1}, \ \boldsymbol{x}_{t+256} \right). \quad (4)$$

### B. Spatial and temporal parallelism

In this research, we focus on iterative stream computation, which is usually seen in time-marching simulation to repeatedly perform the same stream computation for time integral. For iterative stream computation, we can exploit both spatial and temporal parallelism by using processing elements (PEs) with multiple pipelines. Here we assume that a PE updates the entire data for a single time step by streaming them. Fig.2a is a processing element of stream computation, where we use the internal buffer for offset references of streamed data. In this case where only a single PE with a single pipeline is used, no coarse grain parallelism is exploited while fine grain parallelism is available with operators in the pipeline.

By duplicating the pipeline inside the PE as shown in Fig.2b, we can exploit spatial parallelism, or data parallelism to speed up computation for a single time step. Here we share the buffer with the $n$ pipelines to restrict the increase of the buffer size. We fuse independent buffers into a single buffer with multiple inputs and outputs, so that most of the internal memories can be shared. When there is no dependency among computations of data stream elements, we can utilize this spatial parallelism to increase the performance with a similar size of a buffer. However, this approach requires more memory bandwidth due to the $n$ times wider data stream.

To the contrary, we can keep the same bandwidth requirement in increasing the performance by temporal parallelism. As shown in Fig.2c, we can cascade $m$ PEs and use them as a longer pipeline to speed up computation for $m$ time steps. The cascaded PEs require no wider bandwidth to stream data with an external memory because memory accesses are made only at the top and the bottom of the pipeline. Since streaming $T$ data elements through $d$ pipeline stages takes $(T+d)$ cycles, $m$-cascaded PEs take $(T+md)$ cycles while a single PE takes $m(T+d)$ cycles for computing $m$ time steps. When $T >> d$, $m$ times faster computation is achieved by cascading $m$ PEs.

However, this approach has two inherent drawbacks. First, the total buffer size increases. Cascade connection of $m$ independent PEs consumes $m$ times more memories for their internal buffers. Accordingly, on-chip memory resources can limit the number of PEs cascaded when each buffer is large. Second, the utilization of PEs becomes lower due to the prologue and epilogue effects of a pipeline. In pipelining, the performance gain comes from parallel processing with different pipeline stages. Accordingly, some PEs are idle until all the PEs receive data elements to compute at the beginning of pipelining, and after PEs finish computation of the last element. The total effective performance can be much degraded when a short stream goes through a long pipeline.

We can apply both the temporal and spatial parallelism by cascading $m$ PEs with $n$ internal pipelines, giving a design space for various combinations of $(n, m)$. On-chip hardware resources constrain available combinations while their performance depends on several factors including an external memory bandwidth, the depth of pipelines, and the size of stream data. Therefore, given a computing application and an FPGA board, we have to find the one for the best performance and power consumption among available combinations of $(n, m)$.

### C. Stream processing description (SPD)

It is not an easy task to explore design space by designing and implementing various hardware structures in RTL. To improve productivity of design space exploration, we propose a domain-specific language (DSL) for abstracted description of stream-computing hardware. We name the DSL "stream processing description", or SPD. We design SPD for the two major requirements. The first one is to easily and intuitively describe computations with formula, like software codes. The second one is to describe hardware structures in a simple way.

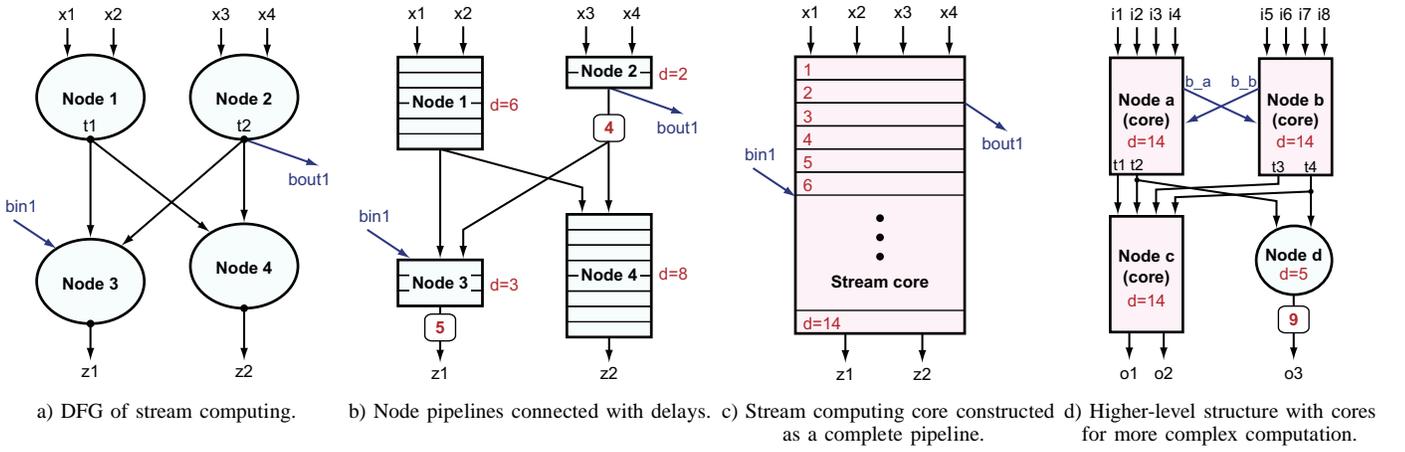

a) DFG of stream computing.  b) Node pipelines connected with delays.  c) Stream computing core constructed as a complete pipeline.  d) Higher-level structure with cores for more complex computation.

Fig. 3.  Hierarchical pipeline construction for stream computing with a data-flow graph (DFG).

```
 1: Name        core;                  # name of this core
 2: Main_In     {main_i::x1,x2,x3,x4}; # main stream in
 3: Main_Out    {main_o::z1,z2};       # main stream out
 4: Brch_In     {brch_i::bin1};        # branch inputs
 5: Brch_Out    {brch_o::bout1};       # branch outputs
 6:
 7: Param       c = 123.456;           # define parameter
 8: EQU   Node1,   t1 = x1 * x2;       # eq (4) (Node1)
 9: EQU   Node2,   t2 = x3 + x4;       # eq (5) (Node2)
10: EQU   Node3,   z1 = t1 - t2 * bin1;# eq (6) (Node3)
11: EQU   Node4,   z2 = t1 / t2 + c;   # eq (7) (Node4)
12: DRCT     (bout1) = (t2);           # port connection
```

Fig. 4.  Stream-processing description (SPD) code for DFG in Fig.3a.

The SPD format allows us to intuitively describe computing formula for pipelines of PEs and connection of the PEs.

In the rest of this section, we use the following example of stream computation with an input vector $\boldsymbol{v}_i^{\text{in}} = (x1, x2, x3, x4)$ and an output vector $\boldsymbol{v}_i^{\text{out}} = (z1, z2)$:

$$t1 = x1 \times x2, \quad (5)$$
$$t2 = x3 + x4, \quad (6)$$
$$z1 = t1 - t2 \times b_{in1}, \quad (7)$$
$$z2 = t1 / t2 + c, \quad (8)$$
$$b_{out1} = t2, \quad (9)$$

where $t1$ and $t2$ are temporary variables, and $c$ is a constant. $b_{in1}$ and $b_{out1}$ are additional input and output streams, respectively. A stream-computing hardware can be implemented as a static mapping of a data-flow graph (DFG) of computation.

*1) Computation description:* Fig.3a shows the DFG of computation for Eqs.(5) to (8), which correspond to Nodes 1 to 4, respectively. Thus, each node represents each formula. The directed edges show the dependences among the formulae. Eq.(9) is an output of Node 2 as bout1. The computation of a formula can be implemented as a pipelined data-path. Fig.3b shows pipelines for the DFG where nodes are replaced with their pipelined data-paths. Since nodes of different formulae can have a different number of pipeline stages, we have to equalize all the path lengths by inserting additional delays.

Figs.4 is an example description in SPD for the DFG of Fig.3a. We describe the computations only with 12 lines. Please note that strings after '#' are treated as comments. Each line is described in a common style of `"Function Fields"` for one of the functions summarized in Table I. These functions are mainly classified into "*core and interfaces*" and "*nodes and connection.*" In the example code of Fig.4, Lines 1 to 5 are for the former. Line 1 names this core with core. Line 2 makes a main stream input interface with a name of main_i and its port names of x1, x2, x3, and x4. Similarly, Line 3 makes a main stream output interface main_o with ports z1 and z2. Lines 4 and 5 make a branch input brch_i with a port bin1, and an output brch_o with a port bout1, respectively.

```
1: Name        Array;
2: Main_In     {main_i::i1,i2,i3,i4,i5,i6,i7,i8};
3: Main_Out    {main_o::o1,o2,o3};
4:
5: HDL Node_a, 14, (t1,t2)(b_a) = core(i1,i2,i3,i4)(b_b);
6: HDL Node_b, 14, (t3,t4)(b_b) = core(i5,i6,i7,i8)(b_a);
7: HDL Node_c, 14, (o1,o2)      = core(t1,t2,t3,t4);
8: EQU Node_d,         o3       = t2 * t4;
```

Fig. 5.  Stream-processing description (SPD) code for the structure in Fig.3d.

The remaining lines are written for nodes and connection. Line 7 beginning with Param defines a parameter cnst with a constant of 123.456, which is used in the formula of Line 11. Such parameters in formulae are statically replaced with their values by a preprocessor. Lines 8 to 11 create Nodes 1 to 4 for a static single assignment to an output port variable with a calculation formula. We refer to this type of a node as *an equation node* or simply *EQU node*. Function EQU is followed by an unique node name and a form of a port variable, '=', and a formula. In a formula, we can use parentheses, operators of +, -, *, and /, and a square root function of sqrt().

In SPD, variables are 32-bit words. For EQU nodes, all related variables are treated as single precision floating-point numbers for numerical computation. When an output variable of a node is referred in a formula of another node, these nodes are connected with a directed edge in the DFG. In addition, we can explicitly connect ports of variables with different names by using DRCT function. Line 12 connects output t2 of Node2 to input bout1 of the branch output interface of Line 5.

*2) Hardware structure description:* The DFG mapped to hardware with pipelined nodes can be considered a single pipeline, as shown in Fig.3c, which can be used as a node in a DFG. This means that we can construct a higher level structure by connecting existing module cores. For example, in Fig.3d, the core of Fig.3c is used as the three nodes connected with another node. Thus we can hierarchically build a hardware structure from a low-level design of data-paths in a core to a high-level design of core connection. This approach provides high productivity in implementing various configurations for parallel stream computation with PEs.

TABLE I. FORMAT OF STREAM PROCESSING DESCRIPTION (SPD).

| Category | Function | Fields | Description |
|---|---|---|---|
| Core and interfaces | Name | \<core name\> | Set a name of this core. |
| | Main_In | {\<IF name\>::port1, port2, ...} | Append input ports for a main stream interface. |
| | Main_Out | {\<IF name\>::port1, port2, ...} | Append output ports for a main stream interface. |
| | Brch_In | {\<IF name\>::port1, port2, ...} | Append input ports for a branch interface. |
| | Brch_Out | {\<IF name\>::port1, port2, ...} | Append output ports for a branch interface. |
| Nodes and connection | EQU | \<node name\>, "equation" | Append an equation node. |
| | HDL | \<node name\>, \<delay\>, "module call", \<param list\> | Append an HDL node. |
| | DRCT | (destination port list) = (source port list) | Connect ports of nodes directly. |
| Others | Param | \= \<constant value\> | Define a parameter with a constant value. |

TABLE II. FORMAT OF SPD SUBFIELD.

| Subfield | Format |
|---|---|
| "equation" | \<output port\> = "calculation formula using input port names" <br> Example: out = ( in1 + in2 * ( t1 - t2 ) ) / in3 + sqrt( in4 ) |
| "module call" | (main output ports)(branch output ports) = \<module name\>(main input ports)(branch input ports) <br> Example: (o1,o2,o3)(do1,bo2) = MyModule(x1,x2,x3,x4,x5)(bi1,bi2,bi3) |

Function HDL is used to create a node with an existing module. We refer to this type of a node as *HDL node*. The pipeline delay of the HDL node has to be statically known in advance of compilation. In HDL line, HDL is followed by a node name, a pipeline delay, module-call description, and a parameter list which is directly passed to parameters of a Verilog-HDL module. The parameter list can be omitted if not necessary. As shown in Table II, "module call" has a similar style to a subroutine call in software, except that multiple output variables can be specified. Fig.5 shows an example of multiple output variables which are written in parentheses. For HDL nodes, all variables are basically treated as raw binary data of 32-bit words while an actual data type in processing depends on each HDL node.

### D. Library modules for HDL node

We can make HDL nodes not only by calling modules described in other SPD codes, but also by using existing modules written in HDL. EQU nodes allow developers to easily describe numerical computation, while HDL nodes extend the function of SPD to arbitrary operations and controls beyond computation. We provide elementary HDL modules in a library that can be used in stream processing without writing Verilog-HDL. The library of the present version contains Synchronous multiplexer, Comparator, Eliminator, Delay, Stream forward, Stream backward, and 2D stencil buffer modules.

### III. DESIGN AND EVALUATION
### A. Overview

As a benchmarking application for DSL-based design space exploration, we chose 2D fluid dynamics simulation based on the lattice Boltzmann method (LBM) [2]. We describe stream computation of LBM in SPD for hierarchical hardware structures; sub-modules for computing stages, a PE consisting of the sub-modules, and cascade connection of the PE. We compile the SPD codes with our stream-computing compiler, which is an extended version of [10], to obtain HDL codes of a custom computing core.

We use an IP-based system integration tool, ALTERA Qsys in order to build a system-on-chip (SoC) common platform consisting of PCI-Express I/F, memory controllers, scatter-gather DMAs, and their interconnects on FPGA. We can easily embed the core generated by the SPD compiler into the system, while this process is not completely automated yet. We also developed a Linux driver and a library software for data transfer between a host program and the FPGA board, and control of stream computation on FPGA.

We compiled the system with the embedded core by using ALTERA Quartus II 14.1 compiler to generate a bitstream for ALTERA Stratix V 5SGXEA7N2 FPGA. We verify FPGA-based fluid dynamics computation with a TERASIC DE5-NET board by comparing the computational results with those by software-based computation. All the designed LBM cores operate at 180 MHz, while 512-bit width DDR3 memory controllers operate at 200 MHz. We evaluate area, active power of the FPGA board, pipeline utilization, and sustained performance per power for design space exploration.

### B. 2D fluid dynamics simulation based on LBM

In the 2D fluid dynamics simulation based on LBM, we compute propagation and collision of fictive particles over a discrete lattice mesh for viscous fluid flow. The details of computation are available in [6]. The computing algorithm of LBM has the three stages of the collision calculation, the translation, and the boundary computation. We wrote SPD codes separately for sub-modules of these stages. Please note that we made three different SPD codes of the translation stage for x1, x2, and x4 parallel pipelines.

Next, we wrote SPD codes to make PEs with $n = 1, 2$, and 4 pipelines. Figs.6 and 8 show the SPD codes for PEs with x1 and x2 pipelines, respectively. Figs.7 and 9 are their compiled DFGs. Here the rounded rectangles are HDL nodes, including the collision calculation node: uLBM_calc, the x1 or x2 translation node: uLBM_Trans2D, and the boundary computation node: uLBM_bndry. The synthesized PEs have 855 and 495 pipeline stages, respectively. Finally, we wrote SPD codes to cascade $m$ PEs. Figs.10 and 11 show the SPD codes for $m = 1$ and 2 cascaded PEs with $n = 1$ pipeline. Fig.12 shows the DFGs of $m = 1$ and 2 PEs with $n = 1$ pipeline. Finally, we implemented six designs for $(n, m) = (1, 1), (1, 2), (1, 4), (2, 1), (2, 2),$ and $(4, 1)$.

```
Name    PEx1;
Main_In  {Mi::if0_0,if1_0,if2_0,if3_0,if4_0,if5_0,if6_0,if7_0,if8_0, iat_0,
          sop,eop, one_tau,rho_in,rho_out};
Main_Out {Mo::of0_0,of1_0,of2_0,of3_0,of4_0,of5_0,of6_0,of7_0,of8_0, oat_0,
          sop,eop};

################ Calculation stage (x1 parallel)
HDL uCalc0, 90,
    (f0_0_c,f1_0_c,f2_0_c,f3_0_c,f4_0_c,f5_0_c,f6_0_c,f7_0_c,f8_0_c) =
    Calc(if0_0,if1_0,if2_0,if3_0,if4_0,if5_0,if6_0,if7_0,if8_0, one_tau);

################ Translation stage (x1 parallel buffer)
HDL uTransx1, 724,
    (f0_0_t,f1_0_t,f2_0_t,f3_0_t,f4_0_t,f5_0_t,f6_0_t,f7_0_t,f8_0_t, at_0_t,
    Mo::sop,Mo::eop) =
    Transx2(f0_0_c,f1_0_c,f2_0_c,f3_0_c,f4_0_c,f5_0_c,f6_0_c,f7_0_c,f8_0_c,iat_0,
            Mi::sop,Mi::eop);

################ Boundary stage (x1 parallel)
HDL uBoundary0, 40,
    (f0_0_b,f1_0_b,f2_0_b,f3_0_b,f4_0_b,f5_0_b,f6_0_b,f7_0_b,f8_0_b,at_0_b) =
    Boundary(f0_0_t,f1_0_t,f2_0_t,f3_0_t,f4_0_t,f5_0_t,f6_0_t,f7_0_t,f8_0_t,at_0_t,
             rho_in,rho_out);

DRCT (of0_0,of1_0,of2_0,of3_0,of4_0,of5_0,of6_0,of7_0,of8_0) =
     (f0_0_b,f1_0_b,f2_0_b,f3_0_b,f4_0_b,f5_0_b,f6_0_b,f7_0_b,f8_0_b);
DRCT (oat_0) = (at_0_b);
```

Fig. 6. SPD code of a stream computing LBM PE with x1 pipeline.

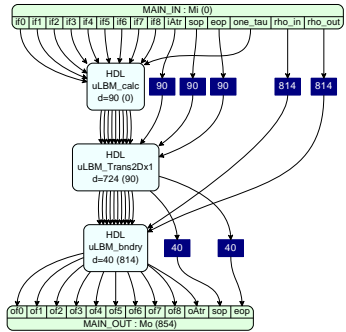

Fig. 7. Stream-computing LBM PE with x1 pipeline.

## C. Resources and performance

Table III summarizes resource consumption of the implemented designs. The SoC peripherals including the PCI-Express I/F and DDR3 memory controllers consume about 23% of ALMs (adaptive logic modules), 6% of on-chip memories, and no DSP block. With the remaining resources, we implemented up to $nm = 4$ pipelines in PEs. Please note that for $nm = 4$, the four cascaded PEs with x1 pipelines consume 3.5 times more on-chip memories than those for the PE with x4 pipelines. This is because the x4 pipelines share a buffer which is slightly larger than the buffer for the x1 pipeline. However, in the case of this 2-dimensional LBM computation, the size of the buffer is very small and negligible. Each pipeline has a total of 131 floating-point (FP) operators in a single precision as shown in Table IV.

Let $N_{\text{Flops}}$ denote the number of FP operators in each pipeline. Since $m$ cascaded PEs with $n$ pipelines perform $nmN_{\text{Flops}}$ operations every cycle once a pipeline is filled, the peak performance is calculated with

$$P(n,m) = nmN_{\text{Flops}}F_{\text{GHz}} \quad [\text{GFlop/s}], \quad (10)$$

where $F_{\text{GHz}}$ is the operating frequency in GHz. In our designs, $F_{\text{GHz}} = 0.18$ and $N_{\text{Flops}} = 131$. Accordingly, the theoretical peak performance is 94.32 GFlop/s for $nm = 4$.

Then we evaluate the utilization of the PE pipelines. By using hardware counters inserted into the top of the LBM computing core, we counted the number of cycles ($n_c$) bringing valid data for computation, and the number of stall cycles ($n_s$) with no computation performed. We calculate the utilization $u$

```
Name    PEx2;
Main_In  {Mi::if0_0,if1_0,if2_0,if3_0,if4_0,if5_0,if6_0,if7_0,if8_0, iat_0,
          if0_1,if1_1,if2_1,if3_1,if4_1,if5_1,if6_1,if7_1,if8_1, iat_1,
          sop,eop, one_tau,rho_in,rho_out};
Main_Out {Mo::of0_0,of1_0,of2_0,of3_0,of4_0,of5_0,of6_0,of7_0,of8_0, oat_0,
          of0_1,of1_1,of2_1,of3_1,of4_1,of5_1,of6_1,of7_1,of8_1, oat_1,
          sop,eop};

################ Calculation stage (x2 parallel)
HDL uCalc0, 90,
    (f0_0_c,f1_0_c,f2_0_c,f3_0_c,f4_0_c,f5_0_c,f6_0_c,f7_0_c,f8_0_c) =
    Calc(if0_0,if1_0,if2_0,if3_0,if4_0,if5_0,if6_0,if7_0,if8_0, one_tau);
HDL uCalc1, 90,
    (f0_1_c,f1_1_c,f2_1_c,f3_1_c,f4_1_c,f5_1_c,f6_1_c,f7_1_c,f8_1_c) =
    Calc(if0_1,if1_1,if2_1,if3_1,if4_1,if5_1,if6_1,if7_1,if8_1, one_tau);

################ Translation stage (x2 parallel buffer)
HDL uTransx2, 364,
    (f0_0_t,f1_0_t,f2_0_t,f3_0_t,f4_0_t,f5_0_t,f6_0_t,f7_0_t,f8_0_t, at_0_t,
    f0_1_t,f1_1_t,f2_1_t,f3_1_t,f4_1_t,f5_1_t,f6_1_t,f7_1_t,f8_1_t, at_1_t,
    Mo::sop,Mo::eop) =
    Transx2(f0_0_c,f1_0_c,f2_0_c,f3_0_c,f4_0_c,f5_0_c,f6_0_c,f7_0_c,f8_0_c,iat_0,
            f0_1_c,f1_1_c,f2_1_c,f3_1_c,f4_1_c,f5_1_c,f6_1_c,f7_1_c,f8_1_c,iat_1,
            Mi::sop,Mi::eop);

################ Boundary stage (x2 parallel)
HDL uBoundary0, 40,
    (f0_0_b,f1_0_b,f2_0_b,f3_0_b,f4_0_b,f5_0_b,f6_0_b,f7_0_b,f8_0_b,at_0_b) =
    Boundary(f0_0_t,f1_0_t,f2_0_t,f3_0_t,f4_0_t,f5_0_t,f6_0_t,f7_0_t,f8_0_t,at_0_t,
             rho_in,rho_out);
HDL uBoundary1, 40,
    (f0_1_b,f1_1_b,f2_1_b,f3_1_b,f4_1_b,f5_1_b,f6_1_b,f7_1_b,f8_1_b,at_1_b) =
    Boundary(f0_1_t,f1_1_t,f2_1_t,f3_1_t,f4_1_t,f5_1_t,f6_1_t,f7_1_t,f8_1_t,at_1_t,
             rho_in,rho_out)();

DRCT (of0_0,of1_0,of2_0,of3_0,of4_0,of5_0,of6_0,of7_0,of8_0) =
     (f0_0_b,f1_0_b,f2_0_b,f3_0_b,f4_0_b,f5_0_b,f6_0_b,f7_0_b,f8_0_b);
DRCT (of0_1,of1_1,of2_1,of3_1,of4_1,of5_1,of6_1,of7_1,of8_1) =
     (f0_1_b,f1_1_b,f2_1_b,f3_1_b,f4_1_b,f5_1_b,f6_1_b,f7_1_b,f8_1_b);
DRCT (oat_0, oat_1) = (at_0_b, at_1_b);
```

Fig. 8. SPD code of a stream computing LBM PE with x2 pipelines.

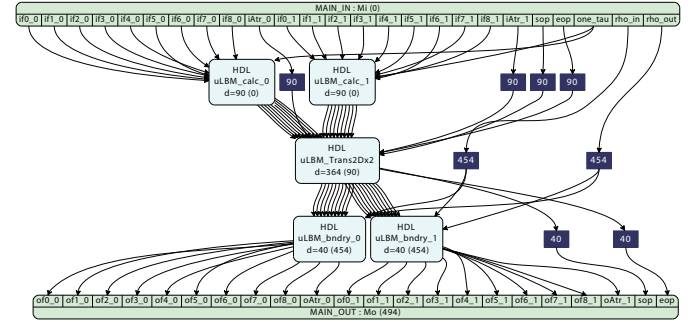

Fig. 9. Stream-computing LBM PE with x2 pipelines.

with $u = n_c/(n_c + n_s)$. As shown in Table III, the utilization is almost 1.0 for PEs with x1 pipeline while PEs with x2 or x4 pipelines have a much less utilization. This is because the DDR3 memory on the FPGA board has only 12.8 GB/s for each of read and write, which can support only the bandwidth required by the x1 pipeline, which is 7.20 GB/s.

By multiplying the utilization and the peak performance, we obtain the sustained performance, which is shown in Table III. In our design space exploration, the configuration of $(n,m) = (1,4)$ gives the best sustained performance of 94.2 GFlop/s, which is very close to the peak. Please note that the negative effect on utilization in pipelining is negligible for a sufficiently large computing-grid, for example, a grid with $720 \times 300$ cells. To evaluate performance per power, we measured the power consumption of the FPGA board by measuring the power supplied by the PCI-Express edge connector with HIOKI power meter PW3336. The highest performance per power is also given by the same configuration of $(n,m) = (1,4)$, which is 2.4 GFlop/sW.

## IV. CONCLUSIONS

This paper presents DSL-based design space exploration to find the best mix of the spatial and temporal parallelism

TABLE III. RESOURCE CONSUMPTION, OPERATING FREQUENCY, PIPELINE UTILIZATION, PERFORMANCE, AND POWER.

| Device / Modules | ALMs | % | Regs | % | BRAM [bits] | % | DSPs | % | Freq. [MHz] | Utilization ($u$) | Performance [GFlop/s] | Power [W] | Perf/W [GFlop/sW] |
|---|---|---|---|---|---|---|---|---|---|---|---|---|---|
| Stratix V 5SGXEA7 | 234720 | 100 | 938880 | 100 | 52428800 | 100 | 256 | 100 | | | | | |
| SoC peripherals | 54997 | 23.4 | 87163 | 9.28 | 3110753 | 5.93 | 0 | 0.0 | - | - | - | - | - |
| ($n$ pipelines, $m$ PEs) = (1, 1) | 34310 | 14.6 | 62145 | 6.62 | 573370 | 1.09 | 48 | 18.8 | 180 | 0.999 | 23.5 | 28.1 | 0.837 |
| (1, 2) | 63687 | 27.1 | 122426 | 13.0 | 1243564 | 2.37 | 96 | 37.5 | | 0.999 | 47.1 | 30.6 | 1.542 |
| (1, 4) | 129738 | 55.3 | 244196 | 26.0 | 2987730 | 5.70 | 192 | 75.0 | | 0.999 | 94.2 | 39.0 | 2.416 |
| (2, 1) | 64119 | 27.3 | 122630 | 13.1 | 642410 | 1.23 | 96 | 37.5 | | 0.557 | 26.3 | 32.3 | 0.812 |
| (2, 2) | 136742 | 58.3 | 244195 | 26.0 | 1316604 | 2.51 | 192 | 75.0 | | 0.558 | 52.6 | 37.4 | 1.405 |
| (4, 1) | 128431 | 54.7 | 243626 | 25.9 | 859604 | 1.64 | 192 | 75.0 | | 0.279 | 26.3 | 33.2 | 0.792 |

```
Name    mQsys_Core10;
Main_In  {Mi::if0_0,if1_0,if2_0,if3_0,if4_0,if5_0,if6_0,if7_0,if8_0,iAtr_0,sop,eop};
Main_Out {Mo::of0_0,of1_0,of2_0,of3_0,of4_0,of5_0,of6_0,of7_0,of8_0,oAtr_0,sop,eop};
Append_Reg {Mi::one_tau, rho_in, rho_out};   ## Definition of constant inputs

################ PEx1_1
HDL Core_1, 495,
    (f0_0_1,f1_0_1,f2_0_1,f3_0_1,f4_0_1,f5_0_1,f6_0_1,f7_0_1,f8_0_1,Atr_0_1,
     sop_1,eop_1) =
    PEx1(if0_0,if1_0,if2_0,if3_0,if4_0,if5_0,if6_0,if7_0,if8_0,iAtr_0,
         Mi::sop,Mi::eop, one_tau,rho_in,rho_out);

DRCT (of0_0, of1_0, of2_0, of3_0, of4_0, of5_0, of6_0, of7_0, of8_0) =
     (f0_0_1,f1_0_1,f2_0_1,f3_0_1,f4_0_1,f5_0_1,f6_0_1,f7_0_1,f8_0_1);
DRCT (oAtr_0, Mo::sop, Mo::eop) = (Atr_0_1, sop_1, eop_1);
```

Fig. 10. SPD code of a single PE with x1 pipeline.

TABLE IV. THE NUMBER OF FLOATING-POINT OPERATORS IN A CORE.

| | Adder | Multiplier | Divider | Total |
|---|---|---|---|---|
| PE with x1 pipeline | 70 | 60 | 1 | **131** |

```
Name    mQsys_Core10;
Main_In  {Mi::if0_0,if1_0,if2_0,if3_0,if4_0,if5_0,if6_0,if7_0,if8_0,iAtr_0,sop,eop};
Main_Out {Mo::of0_0,of1_0,of2_0,of3_0,of4_0,of5_0,of6_0,of7_0,of8_0,oAtr_0,sop,eop};
Append_Reg {Mi::one_tau, rho_in, rho_out};   ## Definition of constant inputs

################ PEx1_1
HDL Core_1, 495,
    (f0_0_1,f1_0_1,f2_0_1,f3_0_1,f4_0_1,f5_0_1,f6_0_1,f7_0_1,f8_0_1,Atr_0_1,
     sop_1,eop_1) =
    PEx1(if0_0,if1_0,if2_0,if3_0,if4_0,if5_0,if6_0,if7_0,if8_0,iAtr_0,
         Mi::sop,Mi::eop, one_tau,rho_in,rho_out);

################ PEx1_2
HDL Core_2, 495,
    (f0_0_2,f1_0_2,f2_0_2,f3_0_2,f4_0_2,f5_0_2,f6_0_2,f7_0_2,f8_0_2,Atr_0_2,
     sop_2,eop_2) =
    PEx1(f0_0_1,f1_0_1,f2_0_1,f3_0_1,f4_0_1,f5_0_1,f6_0_1,f7_0_1,f8_0_1,Atr_0_1,
         sop_1,eop_1, one_tau,rho_in,rho_out);

DRCT (of0_0, of1_0, of2_0, of3_0, of4_0, of5_0, of6_0, of7_0, of8_0) =
     (f0_0_2,f1_0_2,f2_0_2,f3_0_2,f4_0_2,f5_0_2,f6_0_2,f7_0_2,f8_0_2);
DRCT (oAtr_0, Mo::sop, Mo::eop) = (Atr_0_2, sop_2, eop_2);
```

Fig. 11. SPD code of two cascaded PEs with x1 pipeline.

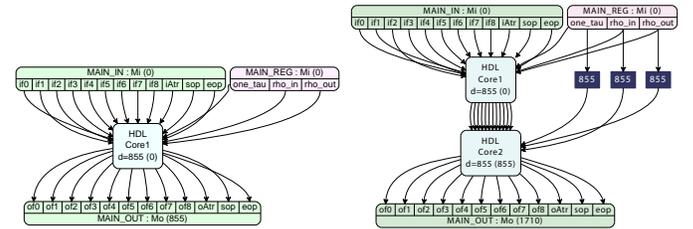

Fig. 12. A single PE and two cascaded PEs with x1 pipeline.

for FPGA-based iterative stream computation. To allow to intuitively describe formulae and submodule calls for various computation and structures of custom hardware in a software-like abstraction level, we propose a domain-specific language, called SPD. Although the SPD compiler is not completely automated yet for the exploration, it allows software developers to design and implement custom hardware with various parallel configurations more easily than doing in conventional RTL languages. Evaluating six configurations of stream-computing cores for fluid dynamics simulation based on LBM, we found the best performance per power is obtained by the design depending only on the temporal parallelism due to the memory bandwidth requirement less than the available one on the used FPGA board.

In the future work, we will automate the process of design space exploration with software codes of a target application.

ACKNOWLEDGMENTS

This research was supported by Grant-in-Aid for Challenging Exploratory Research No.23650021 from the Ministry of Education, Culture, Sports, Science and Technology, Japan.